\documentclass[12pt,preprint]{aastex}
\usepackage{emulateapj5} 
\usepackage{graphicx}
\usepackage{psfig}

\def\ltsima{$\; \buildrel < \over \sim \;$}
\def\lsim{\lower.5ex\hbox{\ltsima}}
\def\gtsima{$\; \buildrel > \over \sim \;$}
\def\gsim{\lower.5ex\hbox{\gtsima}}

\newcommand{\be}{\begin{equation}}
\newcommand{\en}{\end{equation}}
\newcommand{\ergs}{\rm \ erg \; s^{-1}}

\def\cmdue {\rm \ cm^{-2}}

\received{2004 June 21}
\begin{document}
\journalid{}{}
\articleid{}{}

\title{XMM-Newton observation of the double pulsar system J0737--3039} 


\author{Sergio Campana}
\affil{INAF-Osservatorio Astronomico di Brera, Via Bianchi
46, I--23807 Merate (Lc), Italy}
\author{Andrea Possenti, Marta Burgay}
\affil{INAF-Osservatorio
Astronomico di Cagliari, Loc. Poggio dei Pini, Strada 54, I--09012
Capoterra (Ca), Italy}


\begin{abstract}
We report on a 50 ksec XMM-Newton observation of the double
pulsar system J0737--3039 performed on April 2004. We present results
of the spectral analysis of these data combined with
the much shorter Chandra pointing performed on January
2004. Black body emission with effective temperature of $0.20^{+0.02}_{-0.02}$
keV ($90\%$ confidence level) and emission radius $75^{+30}_{-9}$ m for a
distance of 0.5 kpc (implying a 0.5--10 keV luminosity $\sim 6\times
10^{29}\ergs$) is a viable interpretation, calling for a stream of particles
accelerated in the magnetosphere of PSR J0737--3039A and depositing their
kinetic energy in the magnetic polar cap of PSR J0737--3039A or of the 
companion PSR J0737--3039B. A single power-law emission model implies
a very steep photon index $\Gamma=4.2^{+2.1}_{-1.2}$ and a
suspiciously high hydrogen column density, whereas a photon index
$\Gamma=2$ does not provide an adequate description of the XMM-Newton
and Chandra data. A two component model (a black body plus a
power-law with $\Gamma=2$) is statistically acceptable, but the
additional power-law component is not required by the data.
\end{abstract}

\section{Introduction}

PSR~J0737--3039A/B (Burgay et al. 2003; Lyne et al. 2004) is the first
double pulsar ever known. It comprises a 22-ms pulsar
(PSR~J0737--3039A, hereafter A) and a 2.7-sec pulsar
(PSR~J0737--3039B, hereafter B) revolving in 2.4 hr about the common
center of mass along a mildly eccentric ($e=0.09$) and highly
inclined ($\sim 87^\circ$) orbit.  These orbital parameters ensure a
large magnitude for several post-keplerian parameters, making this
binary an unprecedented test-bed for theories of gravity and
relativistic physics (Lyne et al. 2004; Kramer et al. 2004).

Further unique features of J0737--3039 double neutron star system are
the occurrence of both an eclipse of A's radio signal at superior
conjunction (Lyne et al. 2004; Kaspi et al. 2004) and a modulation of
flux and pulse shape of pulsar B along the orbit (Lyne et
al. 2004; Manchester et al. 2004; Ramachandran et al. 2004).  These
phenomenologies are signatures of the interaction between the
relativistic wind from A (releasing a spin-down power
$\dot{E}_A=5.8\times 10^{33}\ergs$, 3600 times larger than
$\dot{E}_B$) and B's magnetosphere and open the possibility of
shedding light on still unanswered issues related with neutron star
magnetosphere, pulsar winds and coherent radio emission (Lyutikov 2004;
Arons et al. 2004; Jenet \& Ransom 2004).

The evidence for strong interactions between relativistic particles
from one pulsar and the other's magnetic field immediately triggered
interest for observations of the J0737--3039 system in the X--ray
band. A short 10 ksec observation with the Chandra ACIS-S
instrument resulted in a
clear detection of the binary (McLaughlin et al. 2004) with no
significant evidence for variability along the orbit on timescale
longer than 15 min. The available 77 photons could be represented
satisfactorily (McLaughlin et al. 2004) by a power-law spectrum with
photon index $\Gamma=2.9\pm0.4$ and $N_H=4.8^{+3.4}_{-2.4}\times
10^{20}$ cm$^{-2}$ ($1\sigma$ confidence error) resulting in a 0.2--10
keV luminosity of $2.3^{+0.5}_{-0.4}\times 10^{30}\ergs$ at the
distance of 0.5 kpc, inferred from the pulsars' dispersion measure and
a model for the Galactic electron density (Cordes \& Lazio 2002). The
inferred X--ray luminosity and spectral index are consistent with an
emission originating solely from the magnetosphere of pulsar A, but
emission from the shocked wind of pulsar A as it interacts with the
interstellar medium (or with the magnetosphere of pulsar B) could be
another viable hypothesis (Granot \& M\'esz\'aros 2004).
 
XMM-Newton targeted J0737--3039 system on April 2004 collecting
about three times more photons than the January 2004 
Chandra pointing. We here report on the analysis of the XMM-Newton data and
present the improved spectral fit made possible 
by joining the data set of the two observations.

\section{XMM-Newton observation}

\subsection{Data collection and reduction}

XMM-Newton observed PSR J0737--3039A/B on 10th April 2004 for 50
ks following an unsolicited target of opportunity request. Data were taken in
Full Frame (full image with 2.6 s timing resolution) and Small Window
($\sim 2'\times 2'$ imaging with 0.3 s timing resolution) mode for MOS1
and MOS2, respectively, and with a medium filter. Data from the pn detector
were taken in timing mode (1 dimensional imaging and 0.03 ms timing
resolution) with a medium filter too. We 
extracted the event files running {\tt emproc} under SAS v6.0. Analysis of the
light curve revealed that a sizable part of the observation was corrupted by
soft photon flares. We extracted, from the entire MOS images, a total light
curve with binning time of 50 seconds and created a time filter
excluding intervals with rate larger than 5 count s$^{-1}$ (e.g. Snowden et
al. 2004). The good
time intervals amount to 20.5 ks and 25.4 ks for the two MOS cameras,
respectively.  In the central part of the MOS1 filtered image we can
detect (using XIMAGE v4.2) only one source. Its location is fully
consistent with that of PSR J0737--3039A/B derived from timing
observation (Lyne et al. 2004), with a count rate of
$(3.3\pm0.5)\times 10^{-3}$ couns s$^{-1}$.  The source turns out to
be too faint for a detection with the pn in timing mode and these data
were not analysed any further.  From the filtered data we extracted
the MOS source spectrum (circular region centered on source with
$r=20''$ extraction radius) and the background spectrum (for the MOS1
using an annular region centered on source with inner and outer radii
$r_{int}=40''$ and $r_{ext}=80''$ respectively, and for the MOS2 using
two $r=20''$ regions near the right corners of the small window
image). The background accounts for $\sim 80\%$ and $\sim 65\%$ of the total
counts for the MOS1 and MOS2 respectively, which are 81 and 145. We generated
response and ancillary files with the latest CCF library and fit
together the two spectra.  We have also analysed the data collected on
18th January 2004 with the ACIS-S instrument onboard Chandra 
finding similar results to McLaughlin et al. (2004). In the latter
data, the source appears slightly extended (with a $\sim 2\sigma$
statistical significance), but the small number of counts (about 80)
does not allow to derive firm conclusions.

\subsection{Spectral fitting}

We tried several emission models to the joint MOS1, MOS2 and
Chandra spectra rebinning the data to 20 and 15 photons per channel for
the two MOS and the ACIS-S respectively; hence a total of 16
spectral channels, ranging from 0.5 keV up to 10 keV for MOS1 and MOS2 (the
photons with energies less than 0.5 keV have been excluded owing to the 
uncertainties in the calibration of the two MOS cameras) and 0.3--10 keV for
the ACIS-S data (corrected for the degradation in the ACIS-S quantum
efficiency within XSPEC with the {\tt ACISABS} component). The adopted
spectral binning guarantees the proper use of the $\chi^2$ goodness-of-fit
evaluator. The spectra have been modelled and fitted using the XSPEC package
v.11.3 with an absorption component modelled by {\tt TBABS}. 

As reported in Table~1, all the explored single component models
are acceptable from the point of view of the goodness-of-fit. However,
their physical interpretation is less straightforward. We review these
models in the following.

A black body model provides a good fit to the data with a
$\chi^2_{red}=1.6$ (where $\chi^2_{red}$ is the reduced $\chi^2$). Given the
small number of degree of freedom (12) the null hypothesis probabily (n.h.p.)
is $7\%$ (Fig. 1). 
The black body equivalent temperature is $k\,T_{\rm bb}=0.20^{+0.02}_{-0.02}$
keV (i.e. $T=2.5^{+0.04}_{-0.05}\times 10^6$ K; here and in the
following we use $90\%$ confidence interval for  
one interesting parameter, i.e. $\Delta\chi^2=2.71$). We can put only
an upper limit on the column density $N_H<0.4\times
10^{21}\cmdue$. This upper limit is consistent with the neutral
hydrogen absorption measured in other sources with similar celestial
coordinates and having distance $\lsim 1$ kpc, as inferred using the
ISM Column Density Search Tool
\footnote{http://archive.stsci.edu/euve/ism/ismform.html}
and from an estimate from the radio pulsars disperion measures of
$N_H=1.5\times 10^{20}\cmdue$.
The unabsorbed 0.5--10 keV flux is $1.9\times 10^{-14}\ergs\cmdue$,
converting into a luminosity of $5.8\times 10^{29}\ergs$.  For the
reference distance of 0.5 kpc, the efficiency $\eta$ of conversion of
rotational energy loss $\dot{E}_A$ into 0.5--10 keV luminosity is
$\eta\sim 10^{-4}$ and the equivalent black body radius is very small
$R_{\rm bb}=75^{+30}_{-9}$ m. Deriving an emitting radius of few
kilometers requires much larger distances for the source, even when
adopting neutron star atmosphere models. In particular, fixing the
emitting radius to 10 km, the atmospheric emission model {\tt NSA} of
the XSPEC package places the source at 13.8 kpc, more than 20
times farther than the distance inferred according to the dispersion
measure distance indicator.

A thermal bremsstrahlung model is statistically acceptable too
($\chi^2_{red}=1.0$, n.h.p. $42\%$). The equivalent temperature is $k\,T_{\rm
br}=0.56^{+0.18}_{-0.19}$ keV. The column density is $N_H<1.0\times 10^{21}\cmdue$ 
The 0.5--10 keV unabsorbed flux is $3.2\times 10^{-14}\ergs\cmdue$. At 0.5 kpc
this translates into a luminosity of $9.5\times 10^{29}\ergs$. The emission
measure for the adopted distance is $3.6^{+6.6}_{-1.1}\times 10^{53}$
cm$^{-3}$.  Following Grindlay et al. (2002) we can estimate the
plasma density $n\sim 10^{26}\,R^{-3/2}$ cm$^{-3}$ for a
spherical region of radius $R$. Assuming a mean particle density
typical of the interstellar medium, i.e. $n=1$ cm$^{-3}$, the radius
of the emitting region turns out $\sim 0.1$ pc, much larger than the
binary separation.  Finally we note that a fit with a Raymond-Smith
model provides an unacceptable result in the case of an emitting
region of solar metalliticy ($\chi^2_{red}=4.3$). The upper limit on
the abundance is $Z<0.01\,Z_\odot$.

A power-law model ($\chi^2_{red}=0.9$, n.h.p. $54\%$) is characterized by a
photon index $\Gamma=4.2^{+2.1}_{-1.2}.$ Given this steep power-law also the
column density is very high $N_H=1.8^{+1.6}_{-0.7}\times 10^{21}\cmdue,$ 
comparable to the full galactic value ($N_H=4.7\times 10^{21}\cmdue$)
derived from neutral hydrogen measurements (Dickey \& Lockman
1990). The unabsorbed 0.5--10 keV flux amounts to $4.7\times
10^{-14}\ergs\cmdue$. Placing PSR J0737--3039A/B at 0.5 kpc, this
results in a luminosity of $1.4\times 10^{30}\ergs$, corresponding
to an efficiency $\eta\sim 2\times 10^{-4}$. We note that a
$\Gamma=2$ model has a null hypothesis probability of $5\times
10^{-7}$ (see Fig. 2).

A fit for a multi-components model keeping all the parameters free is
not meaningful, owing to the limited total number of available photons
and taking in account the reported statistical significance of the
fits for the single component models.  Given its possible physical
relevance (see below), we here only report on a two components model made
by a black body plus a power-law with photon index fixed to
$\Gamma=2$. Also in this case we obtained a good description of the
data ($\chi^2_{red}=1.2$, n.h.p. $28\%$). The black body
component parameters are similar to those of the single black body
model ($k\,T_{\rm bb}=0.16\pm0.04$ keV, $R_{\rm bb}=90^{+83}_{-28}$
m). The percentage contribution to the total flux ($2.7\times
10^{-14}\ergs\cmdue$) of the power-law component is $\sim 52\%$. 

We also attempted a temporal analysis. We folded the MOS light curves
to the orbital period without revealing any significant variation
accross the orbit with an upper limit of $40\%$ ($3\,\sigma$, see Fig. 3). The
aforementioned failure in detecting the source in the pn detector prevented
any search of modulation at the spin period of each the two pulsars.

\section{Discussion}

We have performed a joint spectral analysis of the XMM-Newton
and Chandra observations of the J0737-3039 binary system,
exploiting a three times larger amount of photons than the 
Chandra observation only (McLaughlin et al. 2004).
 
Black body emission seems a viable model. Given the age of the
system ($\gsim 50$ Myr since the second supernova explosion, Lorimer
et al. 2004) the contribution in the 0.5--10 keV band due to the
cooling of the two neutron stars is negligible, but thermal emission
could be powered by a flow of particles impinging onto the surface of
one (or both the) neutron star(s). Pulsar A is likely to provide the
required energetic budget as the spin-down power of pulsar B is only
$2\times 10^{30}\ergs$ (Lyne et al. 2004). The simplest picture calls
for back-flowing of charged particles accelerated in the A's polar gap
and depositing their kinetic energy in the A's magnetic polar cap
(Cheng \& Ruderman 1980; Arons 1981).  The observed equivalent
emitting radius is tiny, much smaller than the polar cap radius of
pulsar A ($r_{\rm pc}\sim 1$ km), but partially filled polar cap
thermal emission can be invoked, accounting for emitting radii as
small as few tens of meters (Harding \& Muslimov
2002). Alternatively, a much larger cap area results from modeling the
data with a non-uniform temperature of the heated region (Zavlin et
al. 2002). In this picture a sinusoidal X-ray emission at the pulsar A
spin period (22.7 ms) is expected.

In fact, a sizable number of millisecond pulsar (MSPs) with black body
spectra has been already observed in the field and in globular
clusters.  In particular thermal contributions and cap radii as small
as 30--40 m have been invoked in order to fit the X--ray spectra of the
nearby pulsar PSR J0437--4715 (Zavlin et al. 2002) in the galactic field. We
also note that the 
equivalent temperature of the emission from J0737--3039 system (0.2
keV) perfectly matches with the one of the pure black body spectrum of
the MSPs observed in 47 Tuc's with Chandra (even if not based on a source by
source spectral fits, Grindlay et al. 2002). The X--ray luminosities (and
hence the emitting radii) are also comparable, once accounted for the
intrinsic uncertainties in the distance of the J0737--3039 system. This
agreement is somewhat surprising given the much older ages of the MSPs in 47
Tuc ($>$ few $10^9$ yrs, Grindlay et al. 2002) with respect to the extimated
time since formation of PSR J0737--3039A.

Alternatively, it is conceivable that a fraction of the A's
relativistic stream of particles is intercepted and channeled onto
pulsar B polar cap(s) whose radius is $80-90$ m (compatible with the
inferred $R_{\rm bb}$ at 95\% confidence level). This hypothesis
requires that $\sim 3\%$ of the spin-down luminosity (supposed
isotropically released) of pulsar A powers the observed X--rays. A
modulation of the X-ray light curve at the B's spin period (or at a
beating frequency between A's and B's rotational periods) is a
predictable consequence of this model. A more detailed model along these
lines has been presented by Zhang \& Loeb (2004). Deeper exposures in small
window with the pn instrument aboard XMM-Newton might reveal this
modulation or the one at the A's spin period, strongly constraining
the origin of X-ray in this binary.

Bremsstrahlung emission requires an emitting region much larger than
the binary separation and also larger (by a factor of $\sim 100$) than
the radius of the bow shock between the relativistic wind of pulsar A
and the ambient medium for a typical density of 1 atom cm$^{-3}$ (and
relative velocity of 200 km s$^{-1}$, Ransom et al. 2004). This
translates in an angular extension of the emitting region of $\sim 1'$
for a distance of 0.5 kpc, which is excluded by both the Chandra
and XMM-Newton observations. 

Absorbed single power-law emission models have been frequently used
for interpreting the X-ray spectra of millisecond pulsars (e.g. Becker
\& Tr\"umper 1999; Becker \& Aschenbach 2002; Nicastro et al. 2004).
In this hypothesis the X--ray emission from J0737--3039 system would be
ascribed to processes involving particles produced and accelerated in
the magnetosphere of pulsar A. However, no MSP with an X--ray power-law
spectrum as steep as $\Gamma\gsim 3$ is known and a power-law model
with a photon index in the range $2\lsim \Gamma\lsim 2.5$ (the
interval of values of $\Gamma$ observed so far in the millisecond
pulsar population, Becker \& Tr\"umper 1999) does not provide an
adequate description of the XMM-Newton and Chandra data.

A simple power-law spectrum can naturally arise also from particles
acceleration in a shock and this picture is particularly interesting
in the case of the J0737-3039 system, where the relativistic wind from
the recycled pulsar A could generate a bow shock at the interface with
the insterstellar medium (provided the velocity of the binary through
the medium is $\gsim 200$ km s$^{-1}$) or when the wind interacts with
the magnetosphere of pulsar B (Granot \& M\'esz\'aros 2004). On the
other hand, both these mechanisms are believed to produce high-energy
spectra with $2\le \Gamma\le 3$ (Granot \& M\'esz\'aros 2004) and a
power-law index $\Gamma \gsim 3$ (at 90\% confidence) appears difficult
to accomodate.

A power-law spectrum with $\Gamma=2$ can model satisfactorily the
observed spectrum of J0737--3039 system only if it is combined with a
dominant thermal emission: in this case, at a distance of 0.5 kpc, the
upper limit to the power-law contribution is at a level of $\sim
4\times10^{29}\ergs$. The latter luminosity would nicely match the
predictions of Granot \& M\'esz\'aros (2004) for the X--ray emission
from a bow shock on the interstellar medium or surrounding pulsar B
magnetosphere. But also a simple magnetospheric emission from pulsar A
\footnote{It is calculated according to the correlation between the
spin-down power $\dot{E}$ of the recycled pulsars in the galactic field
and their observed X--ray luminosity $L_x$, $L_x\propto\dot{E}^\beta$ with
$\beta=1.37\pm0.10$, Possenti et al. (2002). The scatter around this
correlation is about one order of magnitude.}  
is compatible with that luminosity. In fact, a two-component model
(i.e. a thermal emission plus a non-thermal power-law contribution) has
been adopted for improving the fit of the spectrum of PSR J0437--4715 (Zavlin
et al. 2002), but only a long integration with XMM-Newton can assess if it is
required in the case of the J0737--3039 system.

During the refereeing process we became aware of a similar analysis by
Pellizzoni et al. (2004). Even if they did not include Chandra data in the
analysis they reached similar conlclusions.

\clearpage
\begin{table*}
\caption{Spectral fits to J0737--3039.}
\begin{tabular}{ccccc}
\hline
Model     & Column density  & Ph. Index/  & $\chi^2_{\rm red}$   \\
          &  ($10^{21}\cmdue$) & Temperature & (n.h.p.)      \\
\hline
Power-law      &$1.8^{+1.6}_{-0.7}$&$4.2^{+2.1}_{-1.2}$      & 0.9 (0.54)\\
Power-law      & $<0.1$         & 2 (fixed)            & 4.0 ($1\times 10^{-6}$)\\
Bremsstrahlung & $<1.0$         &$0.56^{+0.18}_{-0.19}$ keV& 1.0 (0.42)\\
Black body     & $<0.4$         &$0.20^{+0.02}_{-0.02}$ keV& 1.6 (0.07)\\
NSA            & $<0.5$         &$6.07^{+0.06}_{-0.16}$    & 1.2 (0.24)\\
Black+power    & $<0.9$         &$0.16^{+0.04}_{-0.03}$ keV& 1.2 (0.28)\\
\hline
\end{tabular}

The degrees of freedom are 12 for all the models, 13 for the power-law
with fixed $\Gamma=2$ and 11 for the last model in the table.
\label{spectral}
\end{table*}

\clearpage
\begin{figure*}
\begin{center}
\psfig{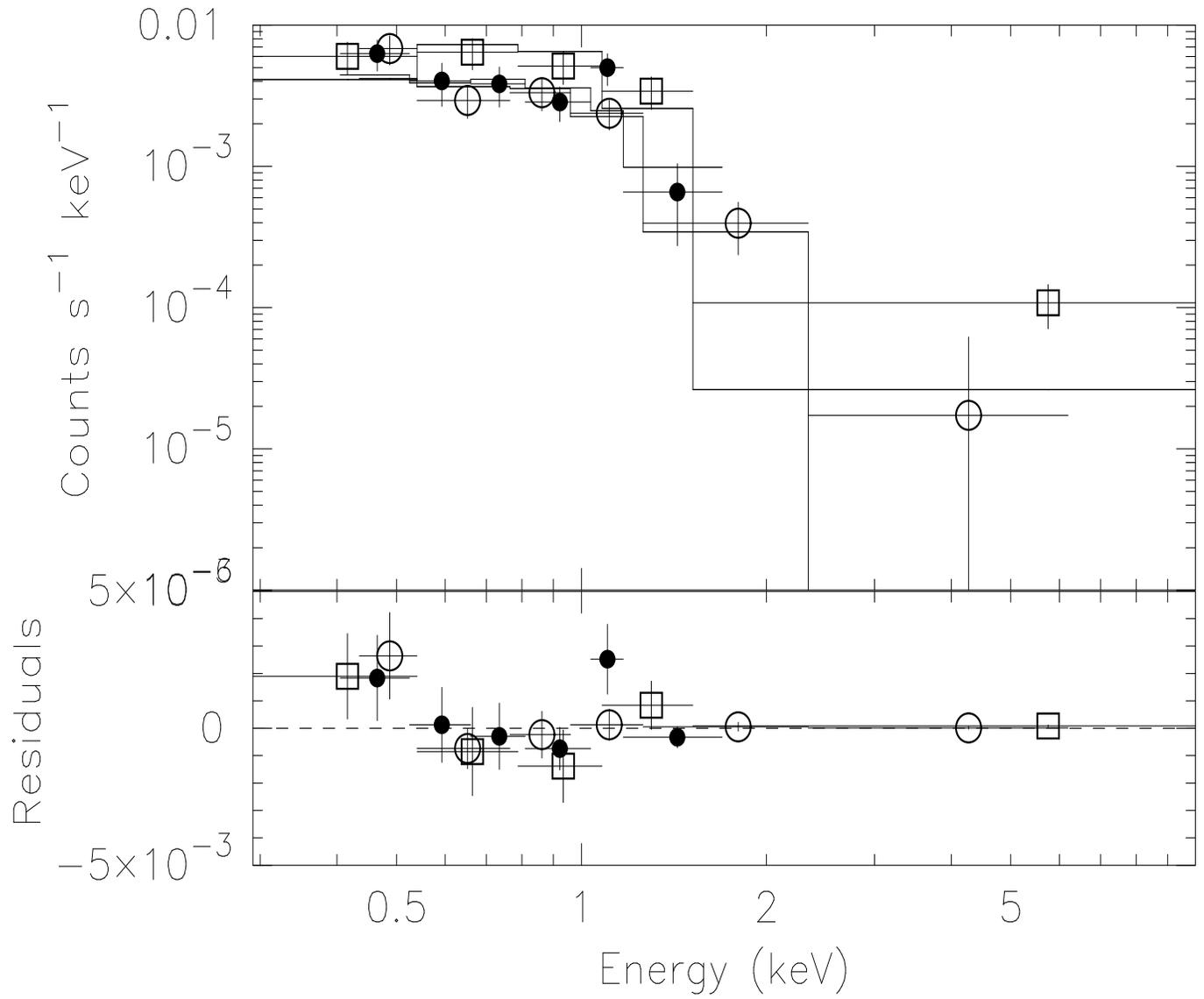}
\caption{Best fit of the pure black body spectrum. Squares indicate Chandra
data, filled dots MOS1 data and open dots MOS2 data.
}
\label{spe}
\end{center}
\end{figure*}

\begin{figure*}
\begin{center}
\psfig{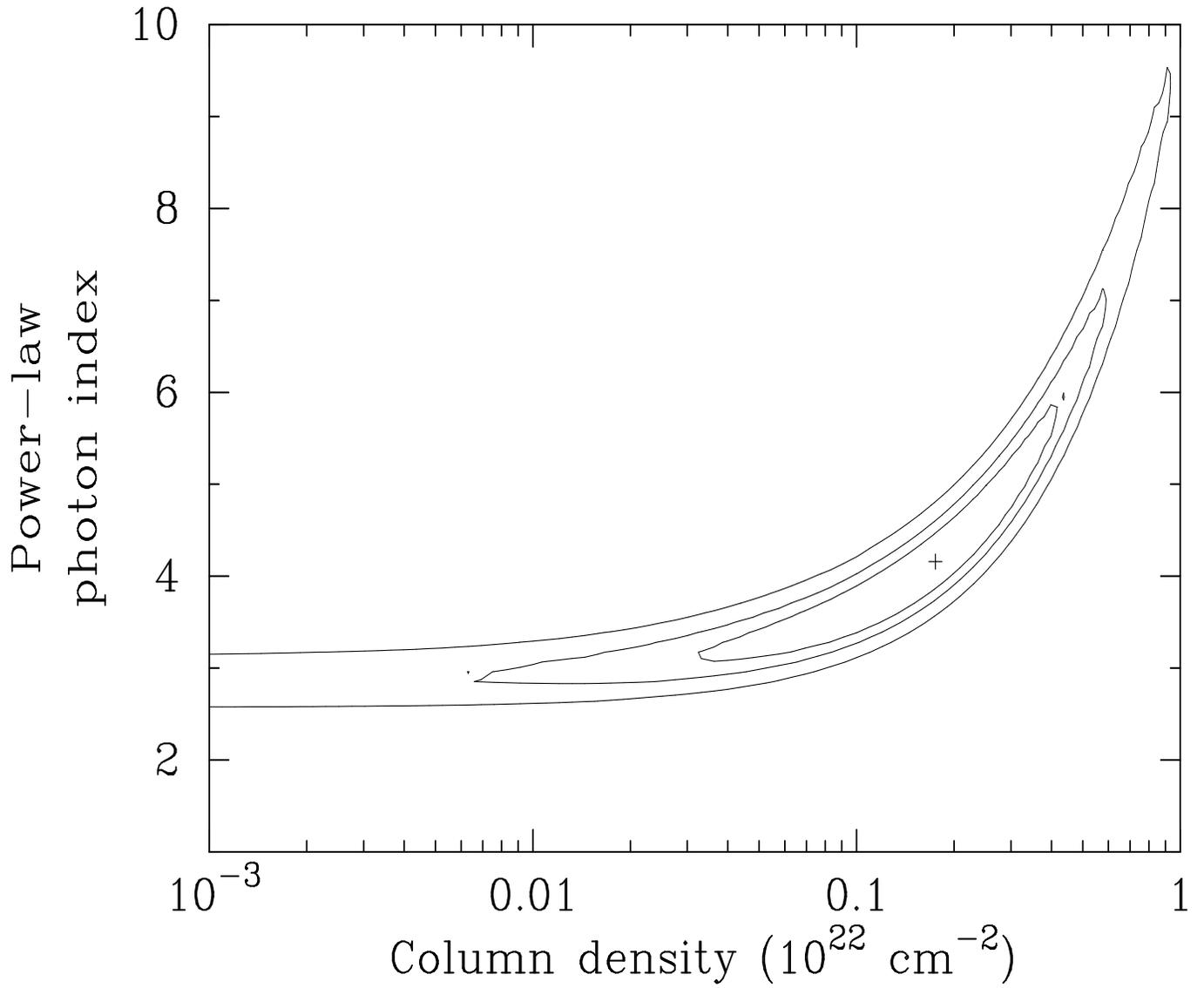}
\caption{Countour plot of the power law fit, ellipses refers to 1, 2 and 3
$\sigma$ levels. A value of the photon index $\Gamma=2$ is outside the
$3\,\sigma$ contour level. Note that the column density axis is in
logaritmic scale.}
\label{cont}
\end{center}
\end{figure*}

\begin{figure*}
\begin{center}
\psfig{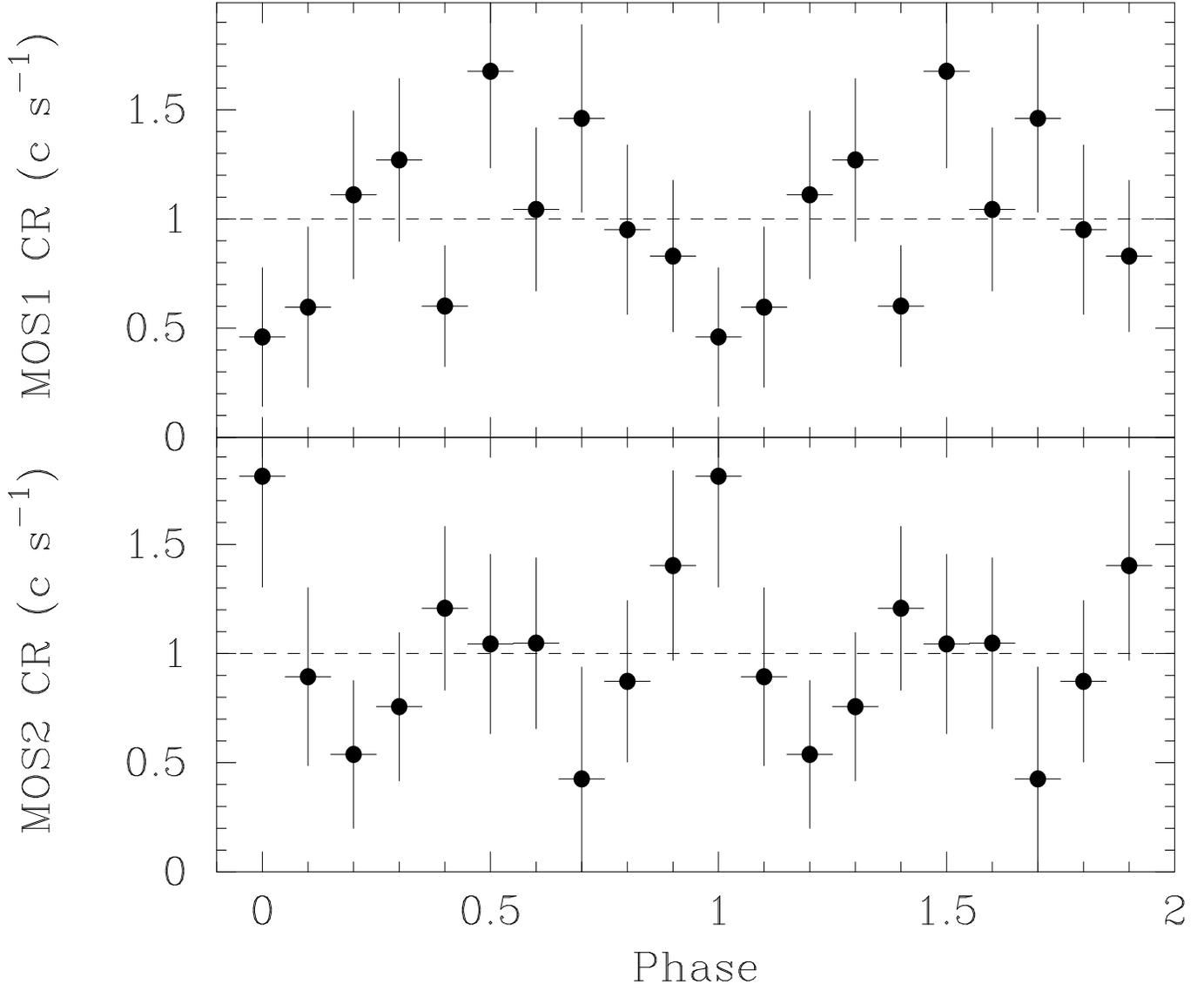}
\caption{MOS1 (upper) and MOS2 (lower) folded background subtracted light
curves. Both curves are consistent with a constant emission.
}
\label{folded}
\end{center}
\end{figure*}

\end{document}